\documentclass[letterpaper,prl,twocolumn,amsmath,amssymb,superscriptaddress,longbibliography]{revtex4-1}
\usepackage[utf8]{inputenc}
\usepackage[draft]{hyperref}
\usepackage{graphicx}% Include figure files
\usepackage{dcolumn}% Align table columns on decimal point
\usepackage{bm}
\usepackage{amssymb}
\usepackage{physics}
\usepackage{color}
\usepackage{soul}
\usepackage{epstopdf}
\usepackage{float}
\usepackage{array}
\begin{document}

\title{Polarisation control for optical nanofibres by imaging through a single lens}

\author{Georgiy Tkachenko}
\email[Corresponding author: ]{georgiy.tkachenko@oist.jp}
\author{Fuchuan Lei}
\author{S\'{i}le Nic Chormaic}
\altaffiliation[Also at ]{Institut N\'eel, Universit\'e Grenoble Alpes, F-38042 Grenoble, France}
\affiliation{Light-Matter Interactions Unit, Okinawa Institute of Science and Technology Graduate University, Onna, Okinawa 904-0495, Japan}

\date{\today}

\begin{abstract}
We present a simple  method for controlling the polarisation state of light at the waist of a single-mode optical nanofibre. The method consists of complete polarisation compensation based on imaging scattered light from inherent inhomogeneities both on the fibre surface and in the glass material itself. In contrast to the recently reported protocol exploiting two imaging systems oriented at 45 degrees to each other, our method requires only one lens and a video camera. It is particularly useful for nanofibre-based applications with severe geometric constraints, such as inside vacuum chambers for  experiments with cold atoms. The measured fidelity of the achieved control is about 98\% using lenses with moderate numerical apertures.
\end{abstract}

\vspace{2pc}
%\noindent{\it Keywords}: nanofibre, polarisation control, scattering

\maketitle

\section{Introduction}
Single-mode optical nanofibres are heavily used in various experiments and applications across classical and quantum optics, atomic physics, and photonics~\cite{tong_oc_2012,nieddu_jo_2016,solano_chapter_2017}. In many cases, it is important to know the polarisation of the guided-mode at the ultrathin waist region. Until very recently, there was no way of measuring or controlling the polarisation at the waist, except for the trivial cases of  horizontal or vertical states, which are identifiable by scattering imaging~\cite{eickhoff_el_1976,goban_prl_2012,vetsch_ieee_2012}. It was impossible to reliably set circularly polarised states, let alone arbitrary elliptically polarised ones, at the nanofibre waist.

%------------------------------------------------
\begin{figure}[h]
\centering
\includegraphics[width=1\linewidth]{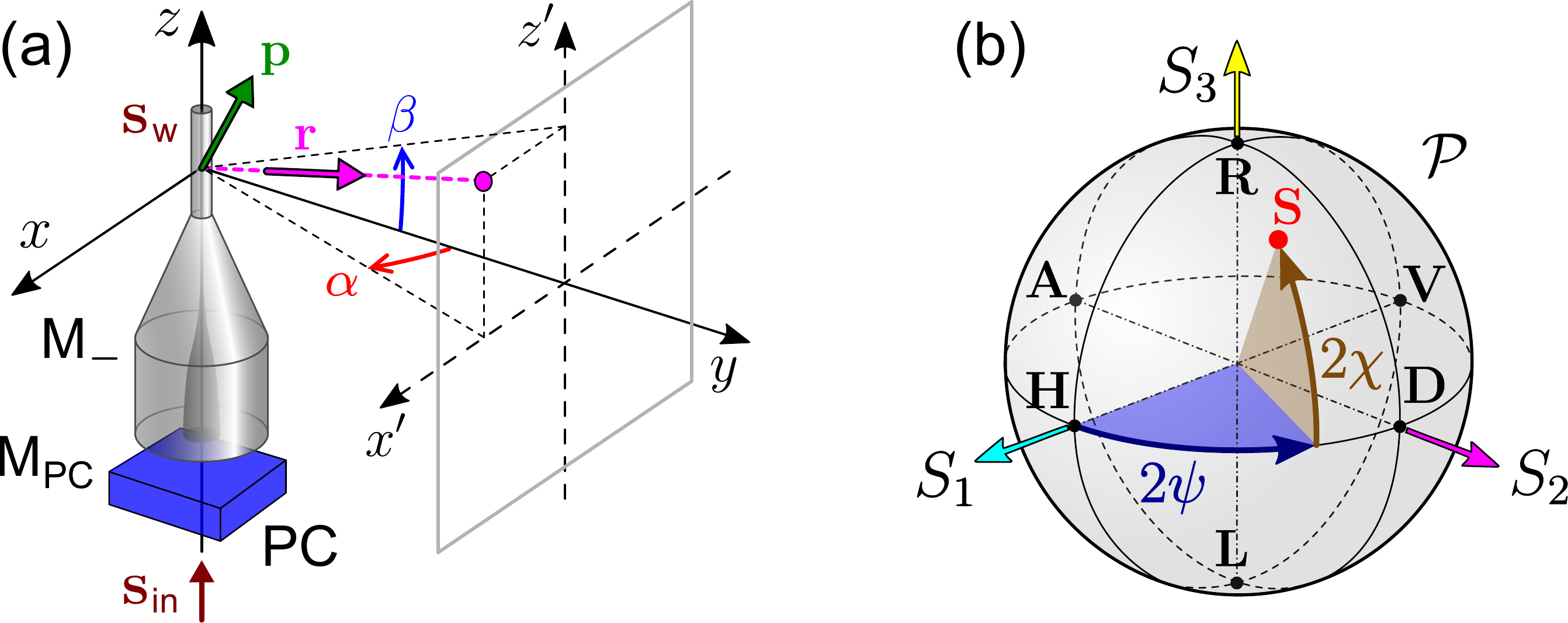}
\caption{(a)~Light escaping from a single-mode nanofibre due to scattering on a point-like inhomogeneity can be defined by the dipole momentum, $\rm p$, which depends on the local polarisation state, ${\bf s}_{\rm w}$. (b)~We describe a polarisation state as a point ${\bf S}$ on the Poincar{\'e} sphere, $\cal P$.}
\label{fig:concept1}
\end{figure}
%------------------------------------------------

Solving this issue, we have recently developed a reliable method for complete control of the polarisation state at the waist of a single-mode nanofibre~\cite{lei_prappl_2019}. The method relies on the fact that, in adiabatically tapered fibres, polarisation transformations are restricted to 3D rotations of the Poincar{\'e} sphere, $\cal P$. Then, by placing a free-space polarisation compensator~(PC) before the fiber, one can reverse any unknown transformation, thereby achieving equality between an arbitrary input state, ${\bf s}_{\rm in}$, and the state at the nanofibre waist, ${\bf s}_{\rm w}$ (see Fig.~\ref{fig:concept1}(a), where the $(x,y,z)$ frame originates at the waist centre and the input beam propagates towards $z>0$).

Light in a completely polarised state is characterised by a Stokes vector ${\bf s} = (1,S_1,S_2,S_3)=(1,\cos2\psi\cos2\chi,\sin2\psi\cos2\chi,\sin2\chi)$, which can be represented on $\cal P$ as a point, $\rm S$, with  angular coordinates $2\psi$ and $2\chi$, see Fig.~\ref{fig:concept1}(b). Using the Stokes-Mueller formalism~\cite{optics_handbook}, the polarisation compensation results in
\begin{equation}
    {\bf s}_{\rm w}={\rm M}_-\,{\rm M}_{\rm PC}\,{\bf s}_{\rm in}={\bf s}_{\rm in}\,,
\end{equation}
where ${\rm M}_{\rm PC}$ and ${\rm M}_-$ are the $4\times4$ real-valued Mueller matrices of the compensator and the down-taper fibre section before the waist, respectively.

The key to our method is to sequentially map two different states from the input to the waist. Importantly, they must be non-orthogonal, i.~e., they must not form an angle of $m\pi$ (where $m$ is an integer) on $\cal P$. In our previous work~\cite{lei_prappl_2019}, this was realised by monitoring the directional coupling of light into a probe nanofibre crossed at right angles to the sample nanofibre. The coupling allowed us to identify the $\bf H$ (or $\bf V$) state and another one, close to $\bf R$ (or $\bf L$). Joos~et~al. have recently achieved polarisation control with a comparable precision, using a different pair of non-orthogonal states: $\bf H$ and $\bf D$~\cite{joos_oe_2019}. These states were identified by monitoring the inhomogeneity-induced scattering from a nanofibre using two imaging systems oriented at $45^{\circ}$ with respect to each other. This approach has an important benefit of being contactless, but its implementation in real experiments with nanofibres may be difficult. In this paper, we report on what may be the simplest realisation of complete polarisation control for single-mode nanofibres, namely, via scattering imaging through a single lens.

\section{Methods}
We assume that the scattering originates from point-like dipolar sources randomly distributed both on the surface and in the bulk of the nanofibre. They are naturally present in optical nanofibres and probably represent nanoscale surface imperfections~\cite{madsen_nl_2016} and bulk material inhomogeneities which do not alter polarisation of the guided light. As we previously found when using a probe nanofibre, the polarisation state, ${\bf s}_{\rm w}$, is maintained throughout the whole waist region~\cite{lei_prappl_2019}.

%------------------------------------------------
\begin{figure}[t!]
\centering
\includegraphics[width=1\linewidth]{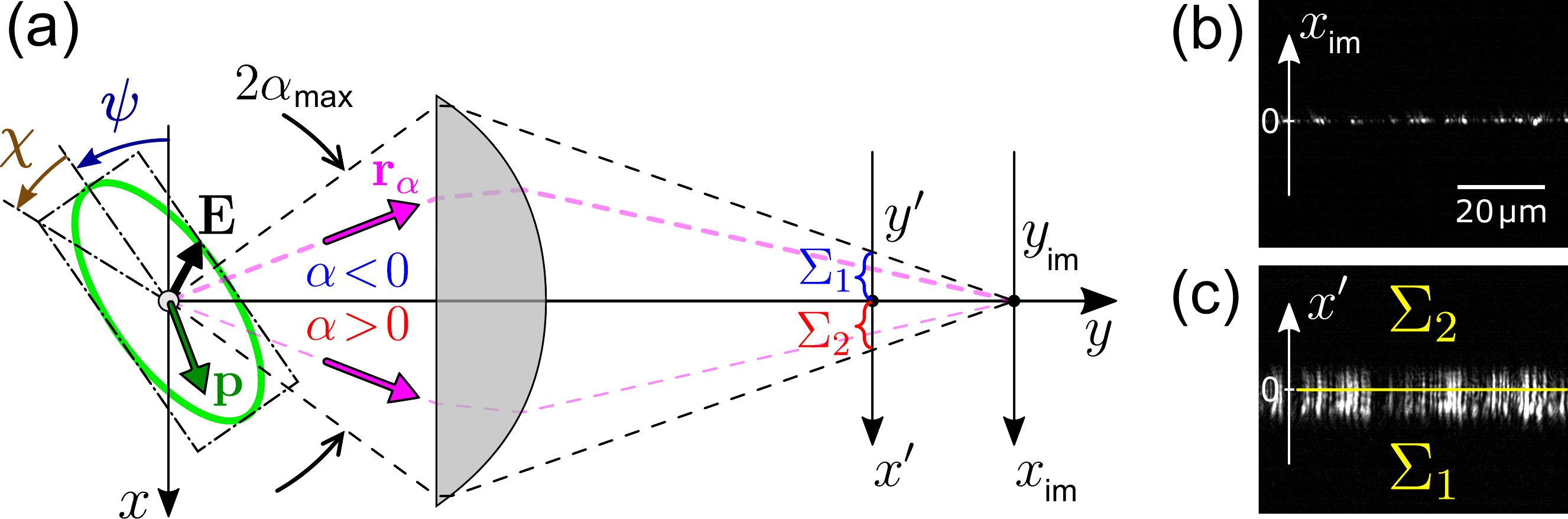}
\caption{(a)~In the $(x,y)$ plane, the power radiated along ${\bf r}_{\alpha}={\bf r}(\alpha,0)$ depends on the polarisation state, ${\bf s}_{\rm w}$. A convex lens collects scattered light within an angular range $|\alpha|<\alpha_{\rm max}$, and allows one to estimate the power difference for $\alpha>0$ and $\alpha<0$. For this purpose, the net power, $\Sigma_1$ and $\Sigma_2$, with $x'<0$ and $x'>0$, respectively, should be measured. (b),(c)~For a point source, $\Sigma_1$ and $\Sigma_2$ cannot be distinguished in the sharp image~(b) located at $y=y_{\rm im}$. Therefore, the blurred image~(c) in the $(x',y')$ plane shall be considered.}
\label{fig:concept2}
\end{figure}
%------------------------------------------------

Let us consider a source located at the coordinate origin and having an electric dipole moment, ${\bf p}$. The time-averaged intensity, $dP$, radiated  into a small solid angle along a radius vector ${\bf r}=(\sin\alpha\cos\beta,\cos\alpha\cos\beta,\sin\beta)$ (with $\alpha$ and $\beta$ being angular coordinates in the $(x,y)$ and $(y,z)$ planes, as defined in Fig.~\ref{fig:concept1}(a)) can be expressed as~\cite{jackson_book}:
\begin{equation}
    dP \propto |({\bf r}\times{\bf p})\times{\bf r}|^2\,.
    \label{eq:dP}
\end{equation}
Unlike free-space beams, guided modes in optical nanofibres may have a significant component of the electric field, ${\rm E}$, along the $z$ axis~\cite{le_kien_oc_2004}. This component, corresponding to $\beta\neq0$,  reduces the brightness difference between scattering images for $\bf H$ and $\bf V$ states~\cite{vetsch_ieee_2012}. Fortunately, this effect can be eliminated by a linear polariser with its axis oriented parallel to $x$. In this study, we place such a polariser in front of the imaging camera, thereby limiting  the detectable scattering to $\beta=0$. The considered dipole momenta are restricted to the $(x,y)$ plane and are linked to the polarisation state as follows:
\begin{equation}
\begin{split}
    {\bf p}(\psi,\chi)& \propto {\sqrt{1+\cos{2\psi}\cos{2\chi}}}\,{\bf e}_x\\
    & +\frac{\sin{2\psi}\cos{2\chi}+i\sin{2\chi}}{\sqrt{1+\cos{2\psi}\cos{2\chi}}}\,{\bf e}_y \,,
\end{split}
\label{eq:p}
\end{equation}
where ${\bf e}_y$ and ${\bf e}_y$ are unit vectors along $x$ and $y$, respectively. In earlier work~\cite{joos_oe_2019}, the $\bf H$ and $\bf D$ polarisations were identified by monitoring the scattering intensities at $\alpha_{\rm H} = 0$ and $\alpha_{\rm D} = 45^{\circ}$, using two imaging systems at these angles. In fact, the same approach was reported earlier in the context of few-mode ultrathin fibres~\cite{fatemi_oe_2015}. In contrast, in the work reported herein, we demonstrate that the angular dependence of $dP$ can be resolved simply by collecting scattered light along various ${\bf r}_{\alpha}$ with a convex lens, as sketched in Fig.~\ref{fig:concept2}(a), where the polarisation state corresponds to the ellipse drawn by the tip of the electric field vector in the $(x,y)$ plane.

A subwavelength nanofibre waist appears in a diffraction-limited sharp image (at $y=y_{\rm im}$) as a line of Airy discs, see Fig.~\ref{fig:concept2}(b), where hardly any difference in brightness for $x_{\rm im}<0$ and $x_{\rm im}>0$ is noticeable when the polarisation varies. In contrast, a blurred image captured in the $(x',y')$ plane (Fig.~\ref{fig:concept2}(c)) shows clearly different brightness sums, $\Sigma_1$ and $\Sigma_2$ (corresponding to $x'<0$ and $x'>0$), for certain polarisation states. The sums can be found by integration of Eq.~\ref{eq:dP} (with $\bf p$ taken from Eq.~\ref{eq:p}) over the relevant angular range: $-\alpha_{\rm max}<\alpha<0$ for $\Sigma_1$ and $0<\alpha<\alpha_{\rm max}$ for $\Sigma_2$, where $\alpha_{\rm max}$ is the maximal half-angle of the cone of light that can enter the lens.

%------------------------------------------------
\begin{figure}[t!]
\centering
\includegraphics[width=0.7\linewidth]{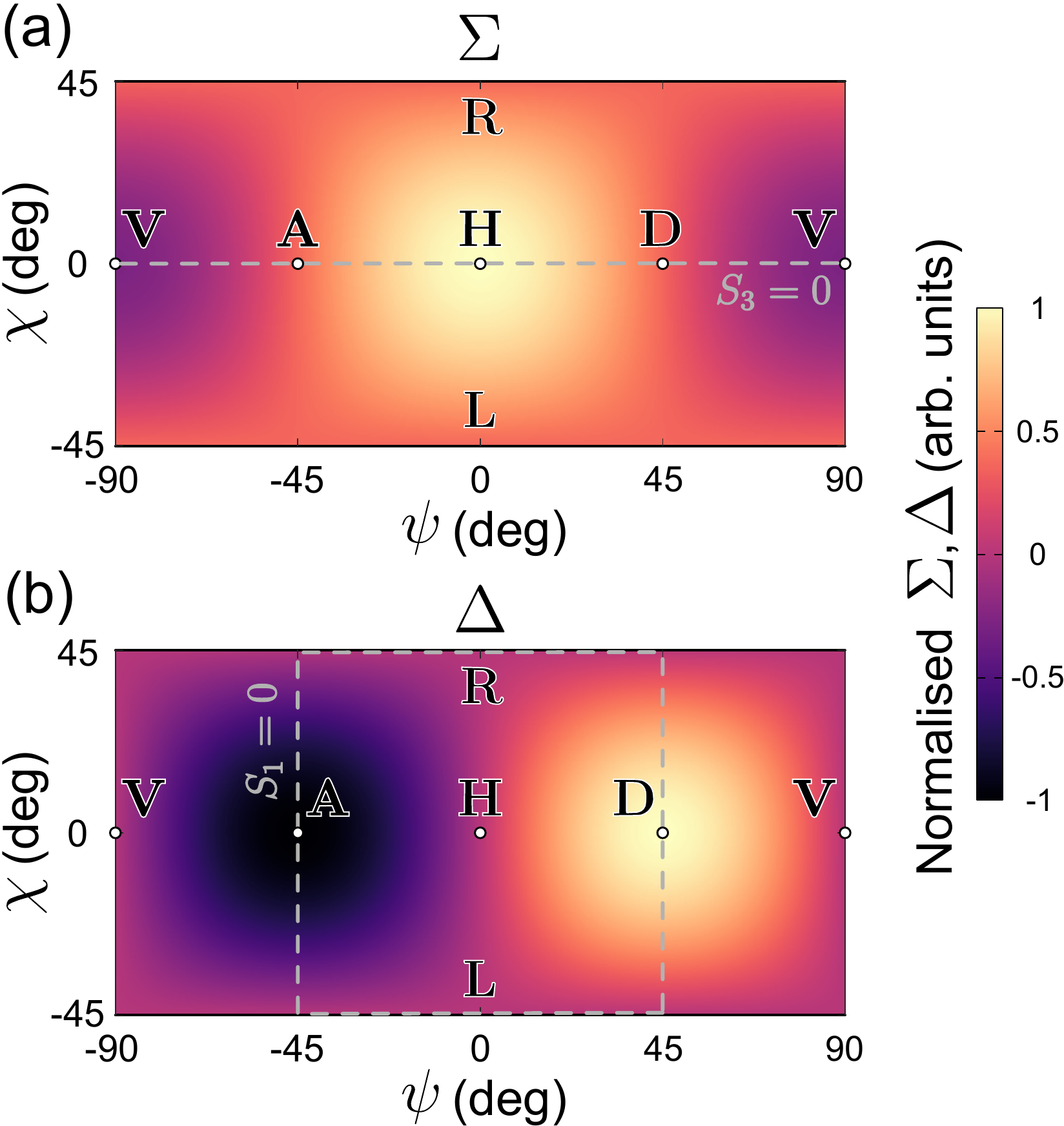}
\caption{Simulated brightness measurements of a scattering image for polarisation states covering the whole $\cal P$ sphere. (a)~The summation signal, $\Sigma = \Sigma_1+\Sigma_2$. (b)~The difference signal, $\Delta=\Sigma_1-\Sigma_2$~(b).}
\label{fig:simulation}
\end{figure}
%------------------------------------------------

Fig.~\ref{fig:simulation} shows the simulation results for $\Sigma = \Sigma_1+\Sigma_2$ and $\Delta=\Sigma_1-\Sigma_2$, normalised by their maxima ($\alpha_{\rm max}=30^{\circ}$ was set). As one can see, $\Sigma$ has the global maximum (minimum) at horizontal (vertical) polarisation, consistent with earlier studies~\cite{goban_prl_2012,vetsch_ieee_2012,joos_oe_2019}. In turn, $\Delta$ has the global maximum (minimum) at a diagonal (antidiagonal) linear polarisation, oriented at $+45^{\circ}$ ($-45^{\circ}$) with respect to $x$. It is worth mentioning that around the $\bf R$ and $\bf L$ states, the spin-orbit coupling in the dipole emission is expected to produce small shifts of the apparent position of the emitter~\cite{araneda_nPhys_2019}. However, these shifts do not exceed $\lambda/(2\pi)$ (with $\lambda$ being the wavelength), and thus are not resolvable in our experimental setup. In addition, the polarisation around the extrema of $\Sigma$ and $\Delta$ is close to linear, in which case the spin-orbit coupling is negligible.

\begin{figure}
\centering
\includegraphics[width=0.7\linewidth]{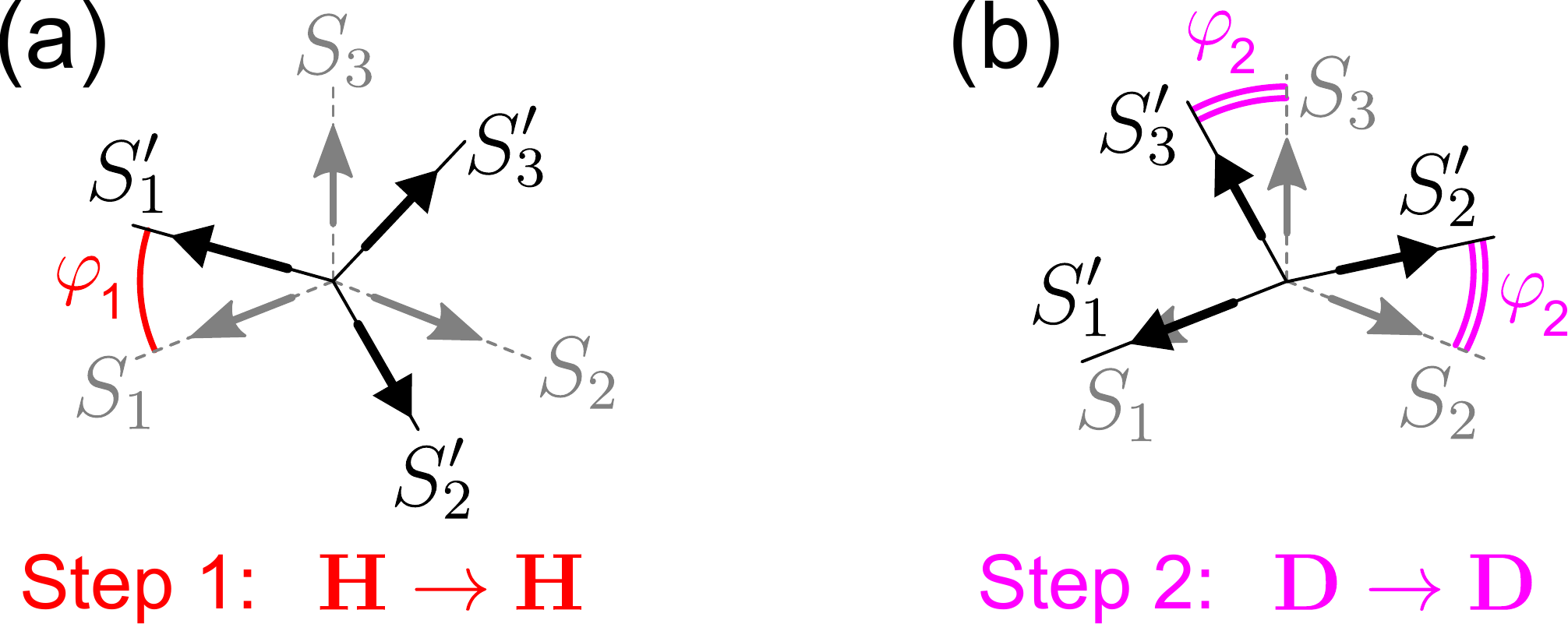}
\caption{Polarisation compensation consists of two steps: (a)~mapping ${\bf H}\to{\bf H}$ to the nanofibre waist by rotating WP1 and WP2 in order to find the maximum of $\Sigma$; (b)~mapping ${\bf D}\to{\bf D}$ by adjusting the retardance of VR to find the maximum of $\Delta$.}
\label{fig:method}
\end{figure}

\begin{figure}
\centering
\includegraphics[width=1\linewidth]{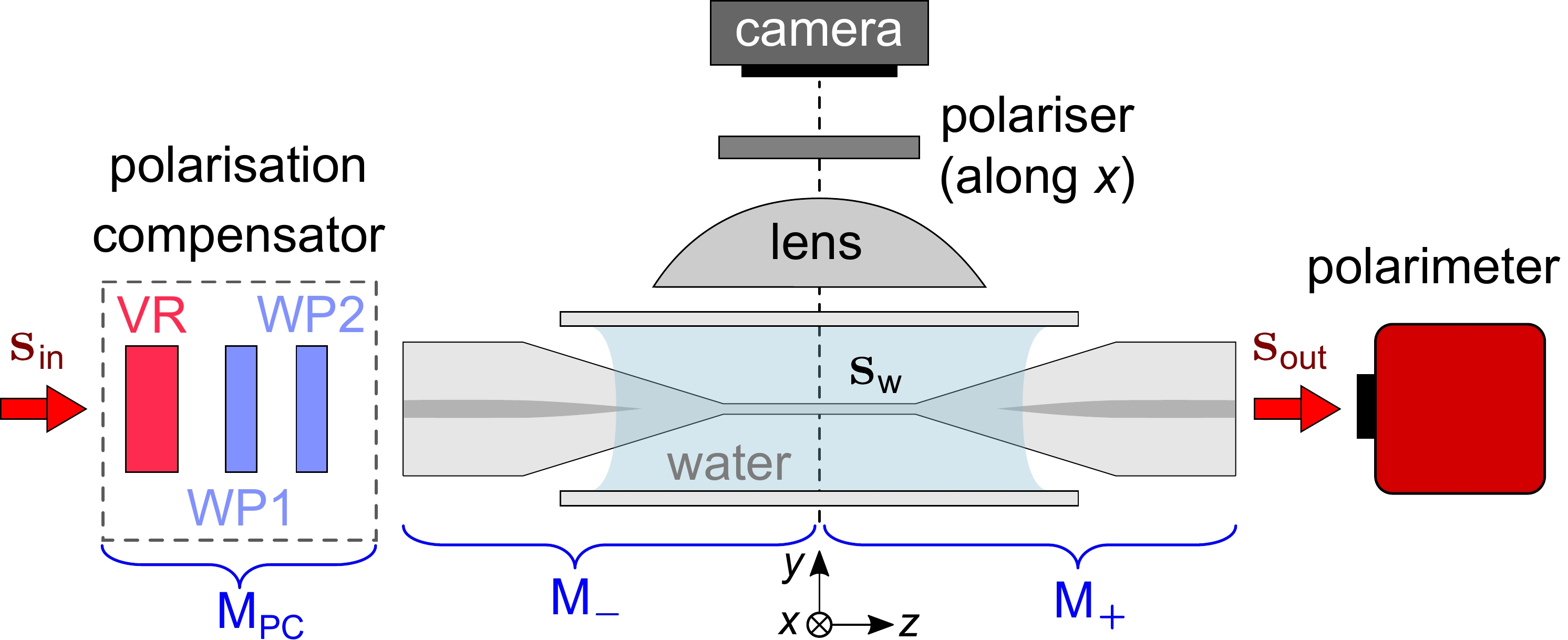}
\caption{The optical setup showing the imaging system and the three-element compensator. Using a two-step compensation procedure based on a scattering image, we achieve ${\rm M}_{\rm PC}={\rm M}_-^{-1}$, and thus ${\bf s}_{\rm w}={\bf s}_{\rm in}$ for any input polarisation.}
\label{fig:setup}
\end{figure}

Based on the simulation results, we perform the polarisation compensation in two steps:
\begin{enumerate}
\item Tilting the $S_1'$ axis of the rotated sphere, ${\cal P}'$, by an angle $\varphi_1$ (Fig.~\ref{fig:method}(a)). This is achieved by setting ${\bf s}_{\rm in}={\bf H}$ and locating the maximum of $\Sigma$ (corresponding to ${\bf s}_{\rm w}={\bf H}$) while allowing ${\bf s}_{\rm w}$ to move freely on ${\cal P}'$ by means of a pair of waveplates, WP1 and WP2 in Fig.~\ref{fig:setup}.
\item Rolling of the $S_2'$ and $S_3'$ axes about $S_1'=S_1$ by an angle $\varphi_2$ (Fig.~\ref{fig:method}(b)). For this purpose, we set ${\bf s}_{\rm in}={\bf D}$ and locate the maximum of $\Delta$ (corresponding to ${\bf s}_{\rm w}={\bf D}$) by adjusting the retardance of a variable retarder, VR, having its optical axis parallel to $x$. Under this condition, ${\bf s}_{\rm w}$ is restricted to the circular trajectory in the $S_1=0$ plane. In the unwrapped $\Delta(\psi,\chi)$ map shown in Fig.~\ref{fig:simulation}(b), this trajectory appears as the dashed square.
\end{enumerate}

%------------------------------------------------
%------------------------------------------------

Our experimental setup is illustrated in Fig.~\ref{fig:setup}. The nanofibre is produced by controlled heating and pulling~\cite{ward_rsi_2014} of a commercial single-mode step-index glass fibre (Thorlabs SM980G80, cut-off vacuum wavelength $0.92\pm0.05\,\mu$m). The cylindrical waist region, of radius $a=0.33\pm0.04\,\mu$m, is connected to taper regions having conical profiles with a half-apex angle of 3~mrad. Such a small angle provides adiabatic coupling between the weakly guided modes of the untapered fibre and the strongly guided ones of the waist~\cite{love_ieee_1991,jung_oe_2008}. We verified the adiabaticity by checking that the fibre maintains at least 97\% transmission at the working vacuum wavelength of $\lambda=1.064\,\mu$m throughout the pulling process. The fibre was coupled to a collimated Gaussian laser beam (Ventus, Laser Quantum Ltd.) with an optical power not exceeding~10~mW. 

The above two-step polarisation compensation procedure is performed by means of~PC, which consists of a pair of quarter-wave plates (WP1 and WP2) and a variable retarder (VR, liquid crystal type, Thorlabs, LCC1111-C). To assess {\it precision} of the polarisation control, we record the output state, ${\bf s}_{\rm out} = {\rm M}_+\,{\bf s}_{\rm w}$, by a free-space polarimeter (Thorlabs, PAX1000IR). The transformation matrix, ${\rm M}_-$, is varied by stressing the input pigtail of the tapered fibre with a three-paddle fibre polarisation controller (FPC, not shown).

%------------------------------------------------
%------------------------------------------------

In order to protect a newly fabricated nanofibre from dust, the waist and taper regions were immersed in a drop of deionized water sandwiched between two parallel glass slides. Under these conditions, we assume that scattering occurs only on point-like, randomly distributed inherent surface or bulk inhomogeneities of the fibre~\cite{vetsch_ieee_2012,joos_oe_2019}. The imaging system, pointed towards $y<0$, comprises a lens, a video camera (Thorlabs, DCC1545M, interfaced with a computer through LabVIEW), and a linear polariser (having its axis parallel to $x$), added in order to maximise the visibility parameter, $V = (\Sigma_{\rm max}-\Sigma_{\rm min})/(\Sigma_{\rm max}+\Sigma_{\rm min})$. We used microscope objective lenses with different numerical apertures (NA): $0.25$ (Olympus PlanN, $10\times$ magnification), $0.55$ (Zeiss LD EC Epiplan-Neofluar, $50\times$), and $1.00$w (water immersion, Zeiss Plan-Apochromat, $63\times$). Depending on the lens, neutral filters with attenuations of $10-20$~dB were placed in front of the camera sensor, to prevent it from saturation. For each of the three tested lenses, we first obtain a sharp image of the nanofibre waist, and then shift the sample towards the camera ($y>0$) by a few percent of the objective's working distance.

\section{Experimental results}
\begin{figure}
\centering
\includegraphics[width=1\linewidth]{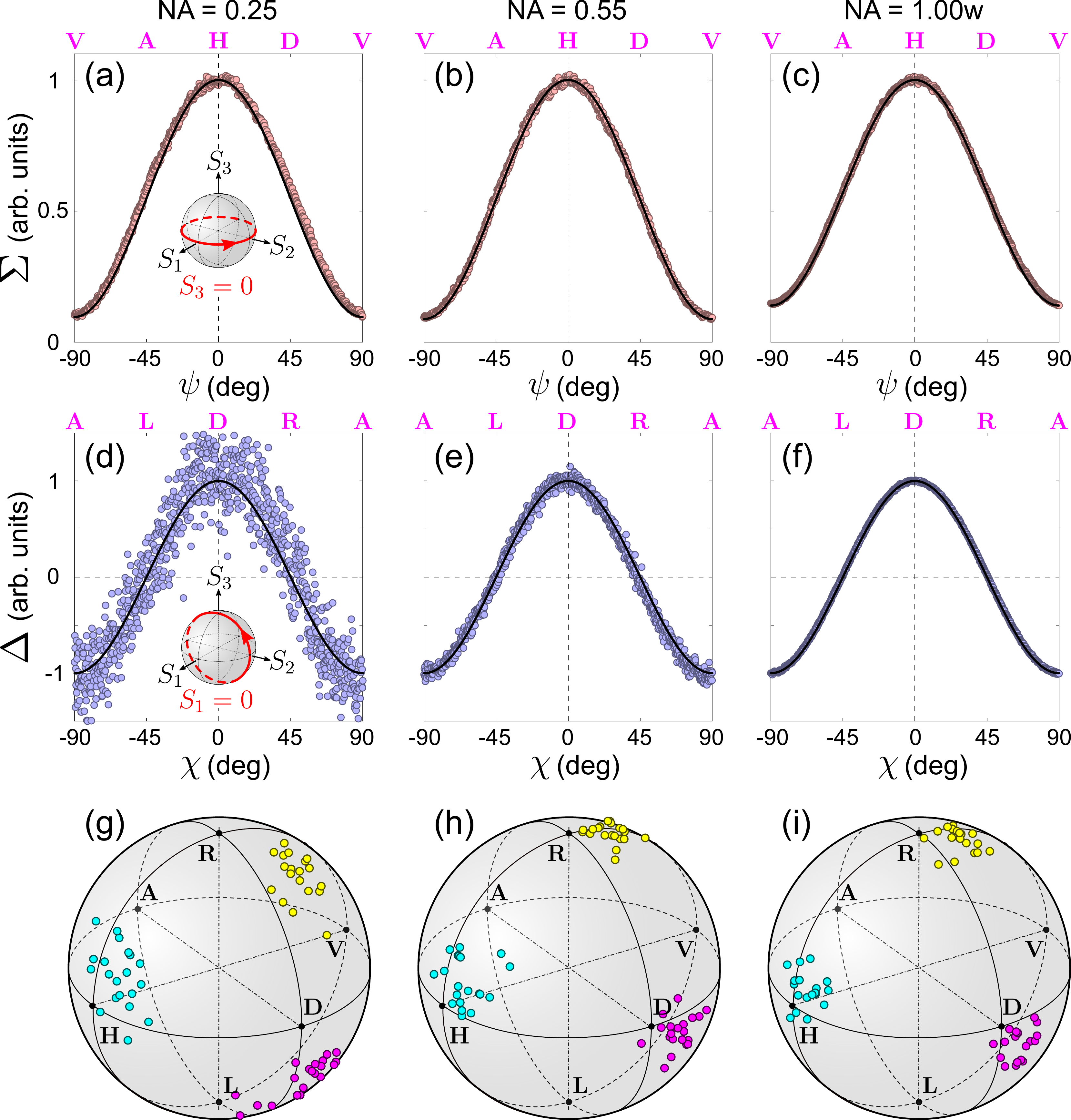}
\caption{(a)-(c)~Normalised summation scattering signal, $\Sigma$, measured with lenses of different NA, versus $\psi$. Inset in~(a): the probed circular trajectory on $\cal P$, corresponding to the dashed line in Fig.~\ref{fig:simulation}(a). Solid lines: fit to the simulated $\Sigma(\psi,0)$, with adjustable~$\alpha_{\rm max}$. (d)-(f)~Normalised difference signal, $\Delta$, versus $\chi$. Inset: the probed circle on $\cal P$, corresponding to the dashed line in Fig.~\ref{fig:simulation}(b). Solid lines: simulated $\Delta(\pm\pi/4,\chi)$. All data shown in (a)-(f) were collected after the polarisation compensation. (g)-(i)~Statistics of the polarisation control assessed by measuring ${\bf s}_{\rm out}$ for ${\bf s}_{\rm in}$ equal to $\bf H$, $\bf D$, or $\bf R$ (cyan, magenta, and yellow markers, respectively).}
\label{fig:results}
\end{figure}

Fig.~\ref{fig:results} summarises the experimental results of this work. After performing the two-step compensation procedure (Fig.~\ref{fig:method}), we tested the simulation results for input polarisation states, ${\bf s}_{\rm in}$, following the circular trajectories at the intersections between $\cal P$ and the plane $S_3=0$ (Figs.~\ref{fig:results}(a)-(c)) or $S_1=0$ (Figs.~\ref{fig:results}(d)-(f)). These cases correspond to tilted linear ($-\pi/2<\psi<\pi/2$, $\chi=0$) and elliptical ($\psi=\pm\pi/4$, $-\pi/2<\chi<\pi/2$) polarisations, with the principal states being labelled above the plots.

%------------------------------------------------
%------------------------------------------------

As evident from Figs.~\ref{fig:results}(a)-(c), $\Sigma$ is virtually independent of NA. By contrast, the noise level in $\Delta$ is very sensitive to NA: the standard deviation of $\Delta(\chi)$ from the model (solid lines in Figs.~\ref{fig:results}(d)-(f)), is approximately proportional to $\alpha_{\rm max}^{-1}$. Interestingly, ${\bf D}\to{\bf D}$ mapping is feasible even with a lens having $\alpha_{\rm max}$ much smaller than $45^{\circ}$, which was the angle between the two imaging directions in earlier studies~\cite{joos_oe_2019,fatemi_oe_2015}.

The statistics of the polarisation control is shown in Figs.~\ref{fig:results}(g)-(i) and summarised in Tab.~\ref{tab1}. For each lens, we performed the polarisation compensation at $20$ random settings of the FPC and measured the output polarisation states, ${\bf s}_{\rm out}$, with the input set at $\bf H$, $\bf D$, or $\bf R$. The obtained clouds of points for $\bf H$ are shifted from the input state by about $10^{\circ}$ on average. This shift appears due to the uncompensated transformation described by the matrix ${\rm M}_+$ which was constant throughout the measurements. The other two principal states, $\bf D$ and $\bf R$, also exhibit a similar shift, which is larger for ${\rm NA}=0.25$ due to the corresponding higher noise in~$\Delta(\chi)$. 

\begin{table}
\centering
\caption{Statistics for the polarisation control with objective lenses of various NA. For each cloud of points on the Poincar{\'e} sphere, we obtained the angular spread, $\hat{\varphi}$ (average deviation from the central point), and the fidelity, $F$ (cosine of $\hat{\varphi}$).}
\begin{tabular}{lrrrrrrrrrrrrrrrr}
            &&\multicolumn{2}{c}{$\bf H$}&&\multicolumn{2}{c}{$\bf D$}&&\multicolumn{2}{c}{$\bf R$}&\\
            \cline{3-4}\cline{6-7}\cline{9-10}
            \multicolumn{1}{l|}{NA} & \multicolumn{1}{c}{$\alpha_{\rm max}$} & \multicolumn{1}{c}{$\hat{\varphi}$} & \multicolumn{1}{c}{$F$} &  & \multicolumn{1}{c}{$\hat{\varphi}$} & \multicolumn{1}{c}{$F$} & & \multicolumn{1}{c}{$\hat{\varphi}$} & \multicolumn{1}{c}{$F$} & & \multicolumn{1}{c}{$V$}\\
            \hline
            \multicolumn{1}{l|}{$0.25$} & $10.8^{\circ}$ & $15.2^{\circ}$ & $0.97$ & & $20.7^{\circ}$ & $0.94$ & & $10.0^{\circ}$ & $0.98$ & & $0.81$\\
            
            \multicolumn{1}{l|}{$0.55$} & $24.2^{\circ}$ & $11.4^{\circ}$ & $0.98$ & & $7.8^{\circ}$ & $0.99$ & & $5.1^{\circ}$ & $0.98$ & & $0.83$\\
            
            \multicolumn{1}{l|}{$1.00$w} & $48.3^{\circ}$ & $7.7^{\circ}$ & $0.99$ & & $11.5^{\circ}$ & $0.98$ & & $11.3^{\circ}$ & $0.98$ & & $0.74$\\
            \hline
        \end{tabular}
\label{tab1}
\end{table}

%------------------------------------------------
\begin{figure}
\centering
\includegraphics[width=0.5\linewidth]{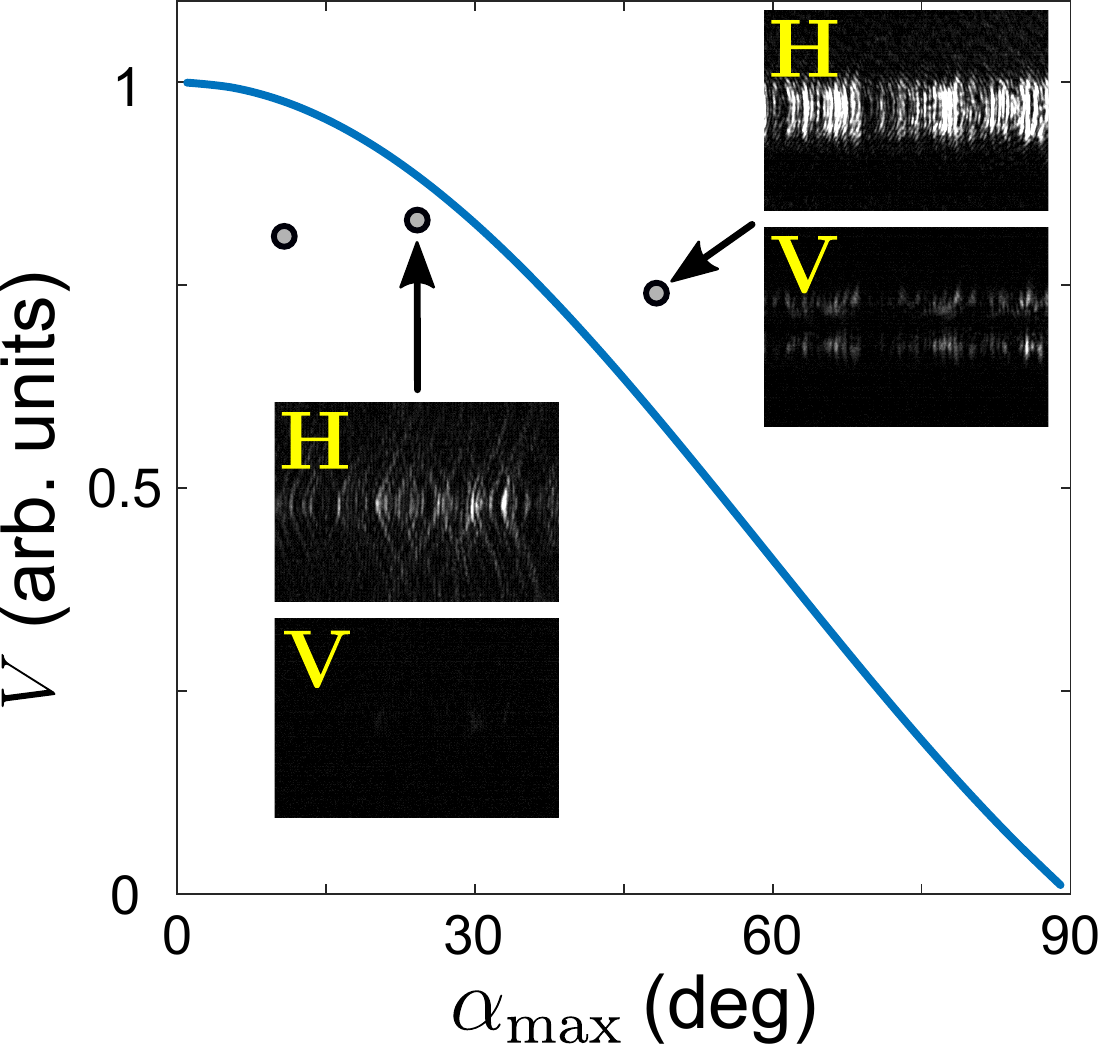}
\caption{Visibility parameter as a function various $\alpha_{\rm max}$. Solid line: simulation; markers: experiment (from Tab.~\ref{tab1}). Insets: scattering images for the $\bf H$ and $\bf V$ states, which correspond to the extrema of~$\Sigma$.}
\label{fig:V}
\end{figure}
%------------------------------------------------

The measured visibility parameter, $V$ (see Tab.~\ref{tab1}), is roughly consistent with the simulations reported in~\cite{joos_oe_2019}. In our case, the linear density of scatterers on the nanofibre waist is about $0.8\,\mu$m$^{-1}$ (as measured with the ${\rm NA}=1.00$w lens). This corresponds to a linear density of $1.23$ per unit of optical resolution taken as $1.22\lambda/(2n\sin\alpha_{\rm max})=0.65\,\mu$m, where $n$ is the refractive index of the medium ($1.34$ for water). According to our simulations, visibility decreases with $\alpha_{\rm max}$, see Fig.~\ref{fig:V}. Indeed, a lens with high NA would capture some light from the sides of the radiation pattern even if the dipole moment were perpendicular to the axis of the imaging system (i.~e., ${\bf s}_{\rm w}={\bf V}$). In order to artificially increase $V$ for the ${\bf H}\to{\bf H}$ mapping step, one can crop the camera image to the paraxial region, effectively reducing~$\alpha_{\rm max}$.

\section{Discussion}
We quantitatively estimate the precision by the measured angular spread, $\hat{\varphi}$ (average deviation from the central point in a cloud), and the corresponding fidelities, $F=\cos\hat{\varphi}$, see  Tab.~\ref{tab1}. These angular spreads are a few times larger than those obtained with the polarisation control using two crossed nanofibres~\cite{lei_prappl_2019}. Yet, the scattering-based compensation performed with higher-NA lenses provides a minimum fidelity of 98\% (the maximum $\hat{\varphi}$ of $11^{\circ}$), which is sufficiently high for many polarisation-sensitive applications. It is worth noting that the measured angular deviations on the Poincar{\'e} sphere are expressed in terms of $2\psi$ and $2\chi$, and, therefore, the corresponding angular deviations of the polarisation ellipse would be half the size. Consequently, our compensation method has about two times smaller errors with respect to the previous study~\cite{joos_oe_2019}, which employed two imaging systems for scattering detection.

This improvement can be attributed to the fact that optical elements in our polarisation compensator have a fixed angle (close to $90^{\circ}$) with respect to the beam axis, $z$. By contrast, operation of the Berek compensator applied in~\cite{joos_oe_2019} involves tilting the optical axis to~$z$. This leads to substantial lateral displacements of the beam (e.~g., up to $150~\mu$m for the model 5540 from Newport$^\circledR$), and, consequently, affects its coupling to the fiber pigtail. As a result, the optical power in the nanofibre depends on the compensator setting. Since the Berek plate is introduced after the ${\bf H}\to{\bf H}$ step, only the ${\bf D}\to{\bf D}$ mapping is affected by variability of coupling efficiency. Still, this may result in a systematic error and diminish the overall {\it accuracy} of the polarisation control.

In fact, our three-element polarisation compensator can be further improved. The pair of quarter-wave plates (WP1, WP2) applied for ${\bf H}\to {\bf H}$ mapping
could be replaced with any set of polarisation optics covering the whole Poincar{\'e} sphere. For instance, a three-paddle FPC from Thorlabs or a single-paddle type from KS~Photonics are good alternatives, as they provide smoother trajectories on the sphere and faster convergence to $\bf H$. For the quickest and most precise mapping, one should consider automatic devices such as the thermal electronic in-line FPC from Phoenix Photonics. In the second step of the compensation, the precision and speed could be enhanced by switching to an automatic scan of the VR retardance, which goal can be readily achieved with an electro-optic modulator interfaced with a computer.

A direct application of the reported method will be in experiments where a single-mode optical nanofibre is integrated into an ensemble of cold atoms in ultrahigh vacuum. Implementation of the method is straightforward: the video camera in the imaging system (commonly used for alignment and characterisation of magneto-optical traps) should be equipped with a linear polariser (as in Fig.~\ref{fig:setup}) and moved slightly closer to the lens in order to achieve a blurred image of the nanofibre, then the two-step compensation procedure can be realised by adjusting a polarisation compensator and monitoring the summation and difference signals from the camera. In ultrahigh vacuum systems, imaging arms typically have ${\rm NA}<0.3$~\cite{alt_opt_2002} corresponding to $\alpha_{\rm max}\approx17^{\circ}$. According to our experimental results (Tab.~\ref{tab1}), even such a modest numerical aperture is expected to provide polarisation control with $95-98$\% fidelity. It could be further improved by using short-focus aspheric lenses, which easily reach ${\rm NA}=0.5$ although one must consider field-of-view constraints for imaging of the atom cloud.

\section{Conclusion}
We have realised complete polarisation control for guided light at the waist of a single-mode optical nanofibre by imaging of scattered light from inherent surface and bulk inhomogeneities of the fibre. We split the image into two parts bordered by the fibre axis and monitor the sum and difference of the net brightness in these parts. This allows us to perform the previously developed two-step polarisation compensation procedure, in a contactless manner, using a simple imaging system comprising a convex lens and a video camera. As a result of the compensation, an arbitrary polarisation state translates from free-space to the nanofibre waist without any change. Our numerical simulations are supported by experimental results obtained for a cylindrical nanofibre waist immersed in water for protection from dust. Statistical studies revealed that lenses with higher numerical apertures provide higher fidelities (equivalently, smaller errors) of the polarisation control. We expect the reported method to be particularly useful for atom-nanofibre hybrid systems where the polarisation of light at the fibre waist is critical. For instance, our method could be applied in novel atom trapping schemes~\cite{phelan_oe_2013,daly_njp_2014}, and for demonstrations of chiral optical forces~\cite{le_kien_pra_chiral_2018} and quadrupole transitions in atomic ensembles~\cite{le_kien_pra_2018}.

\section*{Acknowledgements}
The authors thank J.~M.~Ward and K.~Karlsson for maintaining the nanofibre pulling rig. We also acknowledge S.~P.~Mekhail for assistance with interfacing the video camera. This study was partially supported by the Okinawa Institute of Science and Technology Graduate University. G.~T.~was supported by the Japan Society for the Promotion of Science (JSPS) as an International Research Fellow (Standard, ID~No.P18367).
% \\

\bibliographystyle{iopart-num}
\bibliography{biblio}

\providecommand{\newblock}{}
\begin{thebibliography}{10}
\expandafter\ifx\csname url\endcsname\relax
  \def\url#1{{\tt #1}}\fi
\expandafter\ifx\csname urlprefix\endcsname\relax\def\urlprefix{URL }\fi
\providecommand{\eprint}[2][]{\url{#2}}

\bibitem{tong_oc_2012}
Tong L, Zi F, Guo X and Lou J 2012 {\em Opt. Commun.\/} {\bf 285} 4641--47

\bibitem{nieddu_jo_2016}
Nieddu T, Gokhroo V and {Nic Chormaic} S 2016 {\em J. Opt.\/} {\bf 18} 053001

\bibitem{solano_chapter_2017}
Solano P, Grover J~A, Hoffman J~E, Ravets S, Fatemi F~K, Orozco L~A and Rolston
  S~L 2017 {\em Advances in atomic, molecular, and optical physics\/}
  (Elsevier)

\bibitem{eickhoff_el_1976}
Eickhoff W and Krumpholz O 1976 {\em Electron. Lett.\/} {\bf 12} 405--407

\bibitem{goban_prl_2012}
Goban A, Choi K~S, Alton D~J, Ding D, Lacro\^{u}te C, Pototschnig M, Thiele T,
  Stern N~P and Kimble H~J 2012 {\em Phys. Rev. Lett.\/} {\bf 109} 033603

\bibitem{vetsch_ieee_2012}
Vetsch E, Dawkins S~T, Mitsch R, Reitz D, Schneeweiss P and Rauschenbeutel A
  2012 {\em IEEE J. Sel. Top. Quantum Electron.\/} {\bf 18} 1763--1771

\bibitem{lei_prappl_2019}
Lei F, Tkachenko G, Ward J~M and {Nic Chormaic} S~G 2019 {\em Phys. Rev.
  Appl.\/} {\bf 11}(6) 064041

\bibitem{optics_handbook}
Chipman R~A 2010 {\em Handbook of optics: Ch.~22 Polarimetry\/} (McGraw-Hill)

\bibitem{joos_oe_2019}
Joos M, Bramati A and Glorieux Q 2019 {\em Opt. Express\/} {\bf 27}
  18818--18830

\bibitem{madsen_nl_2016}
Madsen L~S, Baker C, Rubinsztein-Dunlop H and Bowen W~P 2016 {\em Nano Lett.\/}
  {\bf 16} 7333--7337

\bibitem{jackson_book}
Jackson J~D 1999 {\em Classical electrodynamics\/} (Wiley)

\bibitem{le_kien_oc_2004}
{Le~Kien} F, Liang J~Q, Hakuta K and Balykin V~I 2004 {\em Opt. Commun.\/} {\bf
  242} 445--455

\bibitem{fatemi_oe_2015}
Fatemi F and Beadie G 2015 {\em Opt. Express\/} {\bf 23} 3831--3840

\bibitem{araneda_nPhys_2019}
Araneda G, Walser S, Colombe Y, Higginbottom D~B, Volz J, Blatt R and
  Rauschenbeutel A 2019 {\em Nat. Phys.\/} {\bf 15} 17--21

\bibitem{ward_rsi_2014}
Ward J~M, Maimaiti A, Le V~H and {Nic Chormaic} S 2014 {\em Rev. Sci.
  Instrum.\/} {\bf 85} 111501

\bibitem{love_ieee_1991}
Love J~D, Henry W~M, Stewart W~J, Black R~J, Lacroix S and Gonthier F 1991 {\em
  IEE Proceedings J - Optoelectronics\/} {\bf 138} 343--354

\bibitem{jung_oe_2008}
Jung Y, Brambilla G and Richardson D~J 2008 {\em Opt. Express\/} {\bf 16}
  14661--14667

\bibitem{alt_opt_2002}
Alt W 2002 {\em Optik\/} {\bf 113} 142--144

\bibitem{phelan_oe_2013}
Phelan C~F, Hennessy T and Busch T 2013 {\em Opt. Express\/} {\bf 21}
  27093--27101

\bibitem{daly_njp_2014}
Daly M, Truong V~G, Phelan C~F, Deasy K and {Nic Chormaic} S 2014 {\em New J.
  Phys.\/} {\bf 16} 053052

\bibitem{le_kien_pra_chiral_2018}
{Le Kien} F, Hejazi S, Truong V~G, {Nic Chormaic} S and Busch T 2018 {\em Phys.
  Rev. A\/} {\bf 97}(6) 063849

\bibitem{le_kien_pra_2018}
Kien F~L, Ray T, Nieddu T, Busch T and {Nic Chormaic} S 2018 {\em Phys. Rev.
  A\/} {\bf 97} 013821

\end{thebibliography}
\end{document}